# Understanding the Synthetic Pathway to Large Area, High Quality [AgSePh]$_\infty$ Nanocrystal Films

Lorenzo Maserati[†§*], Stefano Pecorario[†‡], Mirko Prato[¶] and Mario Caironi[†*]

[†] Center for Nano Science and Technology @PoliMi, Istituto Italiano di Tecnologia, 20133 Milan, Italy.

[§] The Molecular Foundry, Lawrence Berkeley National Laboratory, Berkeley, CA 94720, USA.

[‡] Micro and Nanostructured Materials Lab, Department of Energy, Politecnico di Milano, 20133 Milan, Italy.

[¶] Materials Characterization Facility, Istituto Italiano di Tecnologia, 16163 Genova, Italy.

**ABSTRACT:** Silver benzeneselenolate [AgSePh]$_\infty$ is a coordination polymer that hosts a hybrid quantum well structure. The recent advancements in the study of its tightly bound excitons (~300 meV) and photoconductive properties (recently employed in UV photodetection) makes it an interesting representative of a material platform that is an environmentally stable alternative to 2D metal halide perovskites in terms of optoelectronic properties. To this aim, several challenges are to be addressed, among which the lack of control over the metal-organic reaction process in the reported synthesis of the [AgSePh]$_\infty$ nanocrystal film (NC). This issue contributed to cast doubts over the origin of its intra-bandgap electronic states. In this article we study all the steps to obtain phase pure [AgSePh]$_\infty$ NC films, from thin silver films through its oxidation and reaction via a chemical vapor-solid with benzeneselenol, by means of UV-vis, XRD, SEM, and AFM. Raman and FTIR spectroscopy are also employed to provide vibrational peaks assignment, for the first time on this polymer. Our analysis supports an acid-base reaction scheme based on an acid attacking the metal oxide precursor, generating water as byproduct of the polymeric synthesis, speeding up the reaction by solvating the PhSeH. The reaction readily goes to completion within 30 min in a supersaturated PhSeH / N$_2$ atmosphere at 90 °C. Our analysis suggests the absence of precursor's leftovers or oxidized species that could contribute to the intra-gap states. By tuning the reaction parameters, we gained control on film morphology to obtain substrate-parallel oriented micro-crystals showing different excitonic absorption intensities. Finally, centimeters size high quality [AgSePh]$_\infty$ NC films could be obtained, enabling exploitation of their optoelectronic properties, such as UV photodetection, in large-area applications.

## Introduction

Low dimensional electronic materials are of wide interest for their strong light-matter interaction,[1] tunable absorption/emission,[2] and for their versatility as building blocks for assembling artificial excitonic devices.[3,4] In the last decade, two-dimensional (2D) transition metal dichalcogenides (TMDs) studies demonstrated a huge potential of this material's platform for both advancing solid state physics understanding and for envisioning new concepts in optoelectronic devices.[5,6] In parallel, 2D hybrid metal-halide perovskites rose as high performance photodetectors and solar cells.[7] Beyond perovskites, covalently bonded confined-in-bulk materials showing higher air-stability were put forward by the supramolecular chemistry community.[8] Among these hybrid quantum wells materials, the metal organic chalcogenide [AgSePh]$_\infty$ has recently emerged as a representative of a potential room-temperature excitonic platform for studying many-body physics,[9] holding at the same time promise in light harvesting and importantly, in UV detection, as we have very recently demonstrated.[10] The silver benzeneselenolate [AgSePh]$_\infty$ is part of a metal-organic chalcogenide (MOC) family where silver and chalcogen atoms form 2D inorganic planes separated by pairs of phenyl rings held together in the out-of-plane direction by Van der Waals forces (see Figure 1a for the extended molecular structure), providing effective quantum confinement along the [001] crystal direction.[9] The first synthesis of [AgSePh]$_\infty$ reported by Cuthbert et al.[11] was obtained by reacting lithium phenylselenolate, generated *in situ* from the reductive cleavage of PhSe-SiMe$_3$ with alkyl lithium reagents, with triphenylphosphine solubilized AgCl, yielding the extended structure [Ag(SePh)]$_\infty$. The crystals were used to refine the crystal structure, but no spectroscopic data were presented (although a "broad absorption" is reported to be peaked at $\lambda = 435$ nm).

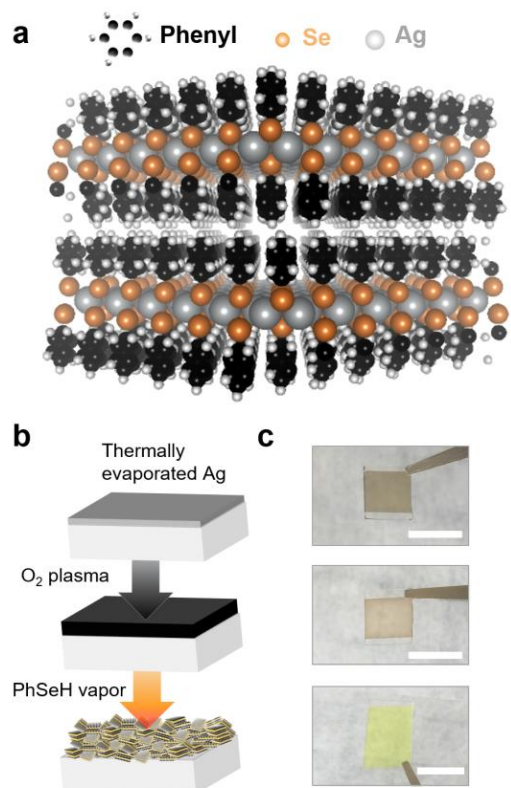

**Figure 1. a,** [AgSePh]$_\infty$ layered nanostructure represented at a molecular level. **b,** Synthetic approach used to yield continuous nanocrystal [AgSePh]$_\infty$ films. **c,** Photographs of the NC films resulting from the corresponding synthetic steps illustrated in **(b)**. Scale bars, 1 cm.

Fifteen years later Hohman and co-workers obtained the same polymer by a completely different route, based on a single-step biphasic immiscible interface method, where silver nitrate and diphenyl diselenide (DPSe) were put in contact at liquid-liquid interface generating micron-size polymer crystals.[12] The same group also introduced, shortly thereafter, a third synthetic pathway to achieve the same material, based on a 3 days in situ acid-base reaction of metallic silver thin films with vapor of DPSe at 80° C.[13] In the same report, they also suggested a similar approach based on bulk Ag$_2$O reacted with liquid benzeneselenol (PhSeH) at 80 °C, also yielding [AgSePh]$_\infty$ micro-crystals. Starting from these works, a first attempt to make continuous large area films to enable spectroscopic experiments has been recently reported by using silver thin films oxidized by UV ozone lamp, subsequently reacted with vapor benzeneselenol.[9] This reaction reportedly yielded square centimeters-size areas covered by nanocrystal films. Nevertheless, the control over the reaction parameters and the understanding of the reaction pathway has been so far missing. Such lack of control prevented also the possibility to study the origin of intra-bandgap states contributing to visible light absorption at energies below the excitonic resonances in such films. Here we provide insights in the aforementioned chemical process and we provide detailed methods for achieving full control over the reaction evolution. By applying vibrational spectroscopies, we unveil unique material's signatures that can help tracing [AgSePh]$_\infty$ growth. We show that silver oxide plays a crucial role, even in its native form on top of metallic silver films, in the crystal nucleation process during the PhSeH-Ag reaction. Indeed, experimental evidences suggest that ultra-thin (<5 nm) silver oxide layers react with PhSeH to yield oriented and separated micron-size crystals. Given the highly anisotropic nature of the excitons, this crystal positioning causes a different absorption line shape due to different exciton polarization in the in-plane versus out-of-plane direction. Instead, thicker silver oxide films exposed to the organoseleniun at high temperature turns into homogenous and continuous and large scale [AgSePh]$_\infty$ nanocrystal (NC) films (Figure 1b,c) that can be used for optoelectronic devices that require charge transport over centimeters squares areas.

## Methods

### Materials

Silicon oxide (200 nm) / doped silicon substrates were purchased from Alpha Nanotech. Alkali-free boro-aluminosilicate glass were purchased from Präzisions Glas & Optik GmbH. Ag 99.99% was purchased from Testbourne LTD. Benzeneselenol 97% was purchased from Sigma.

### Synthesis

*Thermal evaporation.* Corning glass and SiO$_2$/Si$^{++}$ substrates were sonicated for 5 min in acetone, then 5 min in isopropyl alcohol, then blow-dried and O$_2$ plasma for 5 min. The 99.99% silver wire was cut and loaded into a tungsten boat for thermal evaporation, at $2*10^{-6}$ mbar, in a vacuum chamber, inside a glovebox. 5 nm, 20 nm, 50 nm nominal thicknesses were deposited on the chosen substrates with a rate of 0.3 Å/s.

*O$_2$ plasma treatment.* A Diener Electronic Femto Plasma asher was used to oxidize the silver films. Molecular oxygen was injected in the chamber at pressure of 40 mbar with a flux of 0.5 sccm and the plasma with a nominal power of 100 W was regulated with forward and reflected power in terms of percentages. The forward power plasma power was varied from 10% to 100%, keeping the backward zero, with exposure times ranging from 1 to 5 min.

*Metallorganic reaction.* In a nitrogen glovebox, 30 µl of liquid benzeneselenol were pipetted in a Teflon-lined 22 ml vial next to the pre-inserted AgO covered substrate. The sealed vial was transferred in a pre-heated oven at 90 °C. The reaction yielded pure [AgSePh]$_\infty$ phase in about 30 min based on experiments timed from 2 min up to 24 hours. We found important to seal well the vial, not allowing external oxygen to flow-in, possibly oxidizing the benzeneselenol and producing a color change in the reagent toward darker yellow. At the time-stop, the vial was then extracted from the oven and opened under a fume hood. The sample was rinsed in acetone then isopropyl alcohol to remove the unreacted organo-chalcogen reagent, and blow-dried.

### Characterization

*X-ray diffraction.* Spectra were obtained with a BRUKER D8 ADVANCE diffractometer with Bragg-Brentano geometry equipped with a Cu K$\alpha_1$ ($\lambda$ = 1.5440 Å) anode and operating at 40 kV and 40 mA. All the diffraction patterns were collected at room temperature with a grazing angle of 2° and an acquisition range (2$\theta$) between 5° and 80°; Measurements on [AgSePh]$_\infty$ NC films were performed using an angular step of 0.02° with 12 s acquisition time, whereas for the silver oxide analysis a step of 0.05° with 6 s acquisition time.

*UV-vis.* Absorption spectra were obtain by measuring transmission spectra on a PerkinElmer Lambda 1050 UV/Vis/NIR spectrometer.

*Fourier transform infrared spectroscopy.* An FTIR Bruker Vertex 70 was used in transmission mode. Silicon substrate signal was subtracted as baseline reference. The baseline was recorded prior to sample spectra acquisition.

*Raman spectroscopy.* A micro-Raman confocal microscope (inVia Raman Microscope Renishaw) using a 50× objective and an excitation wavelength of 532 nm at 0.037 mW incident power. The detection was optics-limited at 50 cm$^{-1}$.

*Scanning electron microscopy.* Scanning electron microscopy images were collected with a JCM-6010LV, JEOL, with a secondary electron detector and at an accelerating voltage of 2 keV for the electron beam.

*X-ray photoelectron spectroscopy.* Analyzes have been carried out using a Kratos Axis UltraDLD spectrometer. Data have been acquired using a monochromatic Al K$\alpha$ source, operated at 20 mA and 15 kV. High resolution spectra have been acquired at pass energy of 10 eV, energy step of 0.1 eV and take-off angle of 0 degrees with respect to sample normal direction. Analysis area: 300 x 700 microns. Energy scale calibrated on C 1s at 284.8 eV

*Films topography and thickness.* The surface topography of the films was measured with a Keysight 5600LS Atomic Force Microscope operated in the Acoustic Mode. The thicknesses of the films were measured with a KLA Tencor P-17 Surface Profiler.

## Results and Discussion

We make use of a molecular self-assembly method for the synthesis of [AgSePh]$_\infty$ NC films based on the reaction of silver ions with organic benzeneselenol. The optimized reaction parameters for this process are obtained through the exposure of a 40 nm thick AgO film, obtained by mild O$_2$ plasma oxidation of a 20 nm silver film, to a supersaturated (PhSeH)/N$_2$ atmosphere at 90 °C. Considered previous studies and based on the evidence presented hereafter, we describe the reaction in terms of the following chemical equation:

$$4\,\text{AgO}_{(s)} + 4\,\text{PhSeH}_{(l)} \rightarrow 4\,[\text{AgSePh}]_{\infty(s)} + 2\,\text{H}_2\text{O}_{(l)} + \text{O}_{2(g)} \quad (1)$$

The silver oxide phase AgO is a 1:1 molar mixture of silver(I) oxide, $\text{Ag}_2\text{O}$, and silver(III) oxide, $\text{Ag}_2\text{O}_3$.[14,15] It is therefore possible to deconstruct the equation into two steps: first, the mixed phase AgO undergoes thermal decomposition at high temperature to its lower energy state, $\text{Ag}_2\text{O}$. Second, the $\text{Ag}_2\text{O}$ reacts with benzeneselenol to give silver benzeneselenolate. Eq. (1) can be broken down to Eq. (2) and (3):

$$\text{Ag}_2\text{O}_{3(s)} \rightarrow \text{Ag}_2\text{O}_{(s)} + \text{O}_{2(g)} \quad (2)$$

$$2\,\text{Ag}_2\text{O}_{(s)} + 4\,\text{PhSeH}_{(l)} \rightarrow 4\,[\text{AgSePh}]_{\infty(s)} + 2\,\text{H}_2\text{O}_{(l)} \quad (3)$$

Equation (2) is supported by thermal analysis studies performed by Hoflund *et al.* who demonstrated that the kinetically unstable AgO phase decomposes to $\text{Ag}_2\text{O}$ at 100 °C.[5,6] Equation (3) was already put forward[13] as leading acid-base reaction mechanism for converting $\text{Ag}_2\text{O}$ to $[\text{AgSePh}]_{\infty}$. The generation of the reactive $\text{Ag}_2\text{O}$ *in situ* from oxidized Ag by heating in the presence of a benzeneselenol vapor, simplifies our reaction parameters to only: 1) Ag deposition thickness; 2) Ag plasma oxidation extent (power, duration, $\text{O}_2$ pressure); 3) reaction temperature; and 4) reaction time. In comparison to previous methods for synthesizing this hybrid semiconducting crystal, we can produce defect-free, few tens of nanometers-size crystals, covering centimeters squares areas, adequate for the study of its excitonic and transport properties using common techniques. This synthetic method improves upon early generation growth procedures by eliminating variables like water and organoselenide powder reactants to control the rate of the reaction. Our study shows that the morphology and the thickness of the silver and silver oxide films have a crucial impact on the characteristics of the final $[\text{AgSePh}]_{\infty}$ NC film.

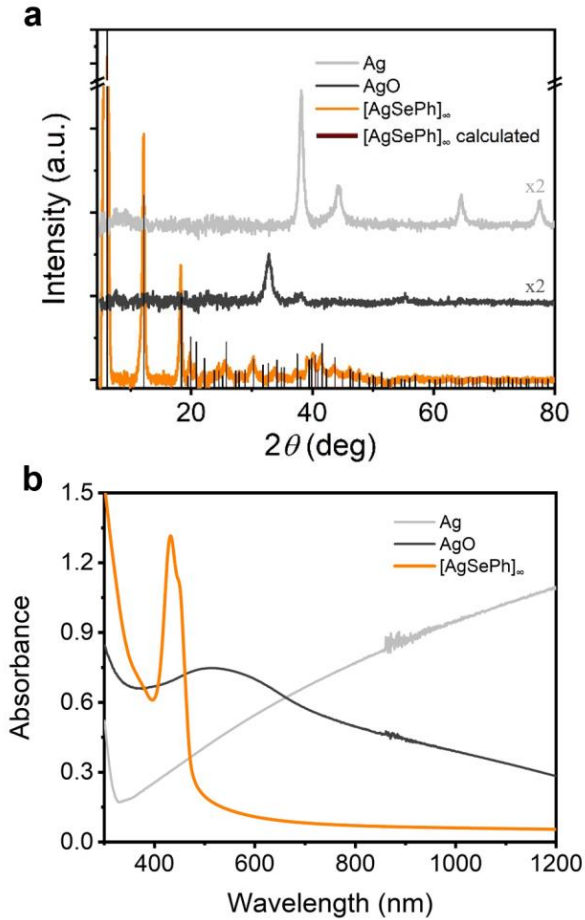

**Figure 2.** X-ray diffraction (XRD) patterns **(a)** and UV-vis spectra **(b)** of Ag (light grey lines), AgO (dark grey lines), and $[\text{AgSePh}]_{\infty}$ (orange lines – experimental, dark orange bars – calculated from ref. 11). The silver was thermally deposited onto glass, it underwent a mild $\text{O}_2$ plasma oxidation and this AgO was reacted with benzeneselenol for 30 min.

X-ray diffraction (XRD) is used to monitor the reaction precursors and products in terms of crystallinity (Figure 2a). The starting 20 nm thick silver film is completely oxidized by a mild $\text{O}_2$ plasma treatment (1 min 10 W, pure $\text{O}_2$ atmosphere at 40 mTorr), and no detectable traces of the silver oxide phase is left after 30 min reaction with PhSeH (the AgO phase is assigned by XRD analysis and corroborated by X-ray photoelectron spectroscopy (XPS) results, Supporting Figure S1). In parallel, UV-Vis spectroscopy (Figure 2b) shows a deep change in the silver optical response after $\text{O}_2$ plasma and a further change upon exposure to PhSeH (reported as $[\text{AgSePh}]_{\infty}$, orange line). To provide a quantitative assessment regarding the products purity we performed XPS on the materials thin films. Having excluded 20% of the oxygen bound to carbon species already observed on the starting Ag film (Figure 3a), we observe silver oxidation upon $\text{O}_2$ plasma leading to 1.4 oxygen to silver ratio (Figure 3b). Following the PhSeH exposure, the oxygen is completely removed from the film (Figure 3c); this observation comes with the *caveat* that the thickness, measured by mechanical profilometry, shows an increase up to 250 nm in the final NC film. Although XPS is a surface sensitive technique, usually not suitable to identify potential "bulk" impurities, the NC film is not compact and we assume that the reported XPS signal fairly represents the atomic content of the analyzed material. By taking a closer look at the XPS peaks (Figure S1), we see that the thermally deposited silver film is indeed composed of Ag(0), with Ag 3d peaks characterized by the asymmetric line shape and peak position in agreement with reports in the literature.[16] Upon mild plasma oxidation, multiple contributions sum up at the silver peaks energies. The best fit is obtained by considering a mixed AgO-$\text{Ag}_2\text{O}$ phase following recent literature insights.[16] The oxygen-silver ratio calculated by integrating the XPS peaks due to silver oxides is close to 0.7, supporting the mixed composition of the surface phase, not observed by XRD because of the different techniques probing depth. We understand this discrepancy in terms of higher environmental stability of $\text{Ag}_2\text{O}$ compared to AgO when exposed to air, causing a surface stabilization of the first.

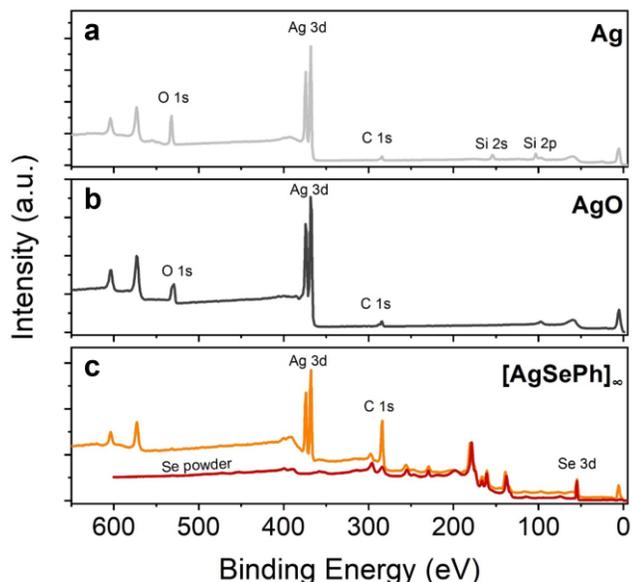

**Figure 3.** X-ray photoelectron spectroscopy (XPS) of Ag (light grey line), AgO (dark grey line), and [AgSePh]$_\infty$ (orange line), together with the spectrum obtained from selenium powder (99.99% purity, Sigma-Aldrich) reference (red line). The main peaks are labelled and used for elemental analysis. The presence of oxygen in the silver sample is due to signal from the SiO$_2$ substrate that vanishes in the double-thickness silver oxide sample and in the much thicker [AgSePh]$_\infty$ NC film. The latest shows no oxygen, and the carbon peak is solely related to phenyls. All the peaks not assigned to Ag, C and O in the [AgSePh]$_\infty$ NC film are clearly due to selenium.

To gain insights in possible metallic silver or oxygen contamination inside the [AgSePh]$_\infty$ NC film, Fourier transform infrared spectroscopy (FTIR) and Raman spectroscopy were applied for the first time on this polymer. For the analysis, thicker samples obtained from 50 nm silver films are reported in the main text for clarity purposes (thinner samples show weaker but analogous peaks, see later discussion on different films thicknesses comparison). Spectroscopic results on the final product are reported in Figure 4 showing the C-C stretches and C-H stretches and bending observed by FTIR (panel a) and Raman (panel b). Highlighted peaks are ascribed to non-centro-symmetric atomic displacements as they are active to both spectroscopies. These peaks are related to the phenyl groups of the benzeneselenol (as it can be seen in Supporting Figure S2) and they dominate the vibrational spectra. The absence of the $\nu$(SeH) stretching mode at 2305 cm$^{-1}$ in the infrared spectrum of [AgSePh]$_\infty$ (Figure 4a) exclude a measurable presence of unreacted/unbound benzeneselenol species.[17,18] In Supporting Figure S2 a direct comparison of the [AgSePh]$_\infty$ NC film with a reported benzeneselenol IR spectra[19] is shown. Moreover, we also note that no detectable Ag or AgO are present: Ag peaks at 444 cm$^{-1}$ and 667 cm$^{-1}$ and AgO peaks at 458 cm$^{-1}$ and 669 cm$^{-1}$ are missing in the [AgSePh]$_\infty$ IR spectrum (Supporting Figure S3a) and the 426 cm$^{-1}$ AgO peak is missing in the Raman spectrum (Supporting Figure S3b).

Then we compare the Raman signatures of [AgSePh]$_\infty$ (Figure 4b) with the surface enhanced Raman scattering data described by K. Kim and coworkers for absorbed benzeneselenol on silver powder.[20] Similarly to what they reported,[20,21] and in agreement with what we have already observed in the IR spectrum, we note the absence of the $\nu$(SeH) stretching mode at 2301 cm$^{-1}$ and the $\delta$(CSeH) scissoring mode at 796 cm$^{-1}$ found in the neat benzeneselenol Raman spectrum. This absence is again indicative of only silver-bounded, deprotonated PhSe species. Then, we use the assignment Table S1 for benzeneselenol monolayer on silver to single out additional peaks present in the [AgSePh]$_\infty$ NC film. We therefore label all the resonances except for three (195 cm$^{-1}$, 97 cm$^{-1}$ and 58 cm$^{-1}$) unrelated to the phenyl vibration or pure Ag-Se vibrations, which are therefore tentatively assigned to intrinsic [AgSePh]$_\infty$ modes. We conclude that no additional molecular species different from [AgSePh]$_\infty$ could be found by means of FTIR and Raman spectroscopy in the analyzed samples. The same claim holds for thinner films that underwent the reaction for one hour or longer under all tested oxidation conditions (see later discussion in the text).

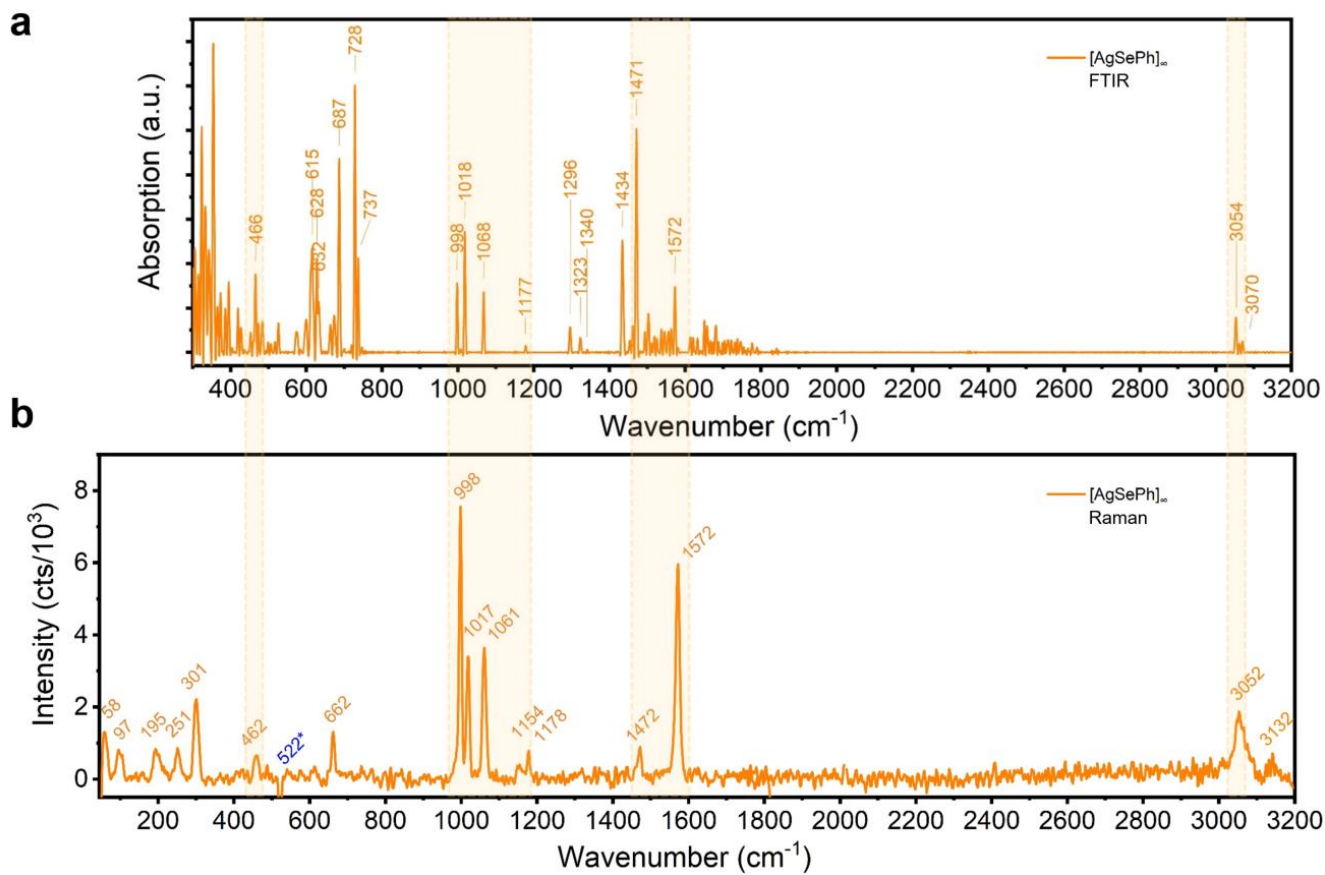

**Figure 4.** Fourier-transform infrared spectroscopy (FTIR) **(a)** and Raman spectroscopy **(b)** plots of [AgSePh]$_\infty$ nanocrystal films. No traces of silver and silver oxide in the [AgSePh]$_\infty$ are found. Shaded areas highlight corresponding peaks in the two spectroscopic techniques, enabled by a non-centrosymmetric unit cell. Distinctive peaks characterize the [AgSePh]$_\infty$ coordination polymer, mainly associated with the phenyl groups (ref. 22, Supporting Table S1) except for the peaks at 58 cm$^{-1}$, 97 cm$^{-1}$ and 195 cm$^{-1}$ assigned to be intrinsic [AgSePh]$_\infty$ modes, and the 522* cm$^{-1}$ peak coming from the silicon substrate. FTIR signal below 450 cm$^{-1}$ is not considered in the analysis due to the instrumental noise, which also affects the measurement in the 1450 – 1800 cm$^{-1}$ range.

Since the oxidation of silver is a critical step in the synthesis of the [AgSePh]$_\infty$ NC film, details regarding O$_2$ plasma reaction are important. Indeed, considering the extreme case, without oxidation, a 20 nm thin metallic silver film promotes the growth of flat, sparsely arranged, micron-scale [AgSePh]$_\infty$ crystals over the 4 h reaction (Figure S4a). The UV-vis spectrum of such sample shows a small excitonic bump (Figure S4c), interestingly different from the exciton spectral shape of Figure 2b. This is an effect of the anisotropic excitonic resonances and indicative of the different in-plane versus out-of-plane contributions, as reported elsewhere.[9] Instead, even just a mild O$_2$ plasma oxidation of the silver film leads to a completely different result (after 30 min PhSeH reaction), characterized by a nanocrystalline morphology and a characteristic excitonic resonance (Figure S4b,d, respectively). Oxygen plasma was chosen for achieving a higher control over the silver oxidation homogeneity compared to the UV ozone lamp used in previous studies.[9] The silver film nominal thickness was held fixed at 20 nm, while different O$_2$ plasma power and times (see Experimental Section) were tested. Figure 5a,b illustrates the evolution of the silver to silver oxide crystallinity (a) and light absorption (b). Upon short plasma exposure, the silver film oxidizes completely leaving no detectable metallic silver. As the oxidation proceeds, the newly formed silver oxide evolves through different crystal phases. While the variation is clearly visible from Figure 5a, the assignment of such phases is more subtle and it is detailed in Supporting Figures S5,6. The tetragonal AgO seems to be the favored crystal phase forming under prolonged O$_2$ plasma exposure, while for mild plasma treatment a mixture of multiple phases is observed. Considering also the XPS results (Supporting Figure S1), a very thin Ag$_2$O layer (not detected by XRD) could be present at the surface after air exposure, as this is the most stable coordination.

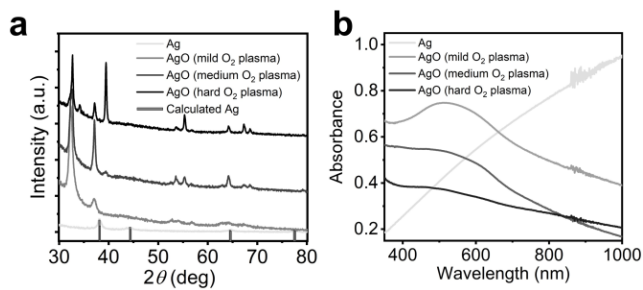

**Figure 5.** X-rays diffraction (XRD) patterns **(a)** and UV-vis spectroscopy **(b)** of Ag (grey line) oxidation into AgO (darker grey lines corresponding to mild, medium and hard oxidation) due to increasingly more aggressive O$_2$ plasma treatments.

Although all different silver oxides phases lead to [AgSePh]$_\infty$ NCs upon benzeneselenol reaction (see Supporting Figure S7), the most crucial aspect of O$_2$ plasma appeared to be the morphology change in the silver film, which in turn influences the morphology at large scale of the [AgSePh]∞ film. In fact, by observing the atomic force microscopy (AFM) images of the silver films before and after different oxidation intensities (Figure 6), a substantial variation of film roughness emerges. Considering a 20 nm thin silver film (Figure 6a), only a mild oxidation preserves a root mean square (RMS) roughness (reported for a 2 x 2 μm area, it does not change significantly up to 10 x 10 μm inspected area) of the same order of magnitude of the starting silver film (~2 nm), while harsher plasma conditions (Figure 6c,d) lead to silver oxide films with much bigger grain size, spanning tens of nanometers (~20 nm to ~40 nm RMS roughness).

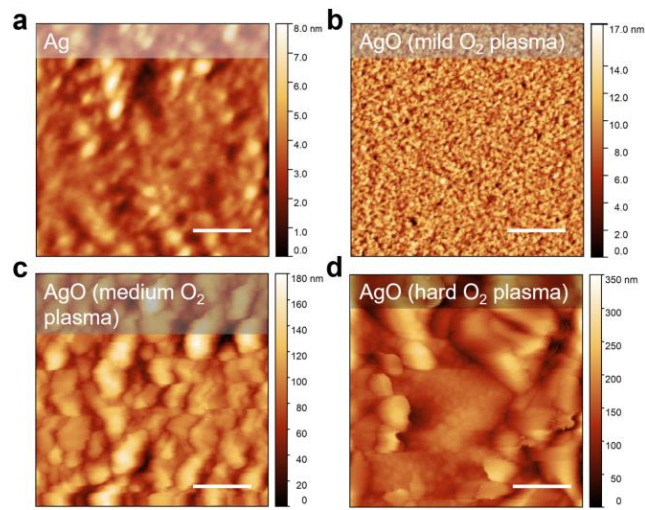

**Figure 6.** Atomic force microscopy (AFM) of a 20 nm silver film, before and after different O₂ plasma treatments. **a,** Ag as thermally deposited. **b,** Ag thin film after 1 min O₂ plasma 10 W (mild treatment). **c,** Ag thin film after 1 min O₂ plasma 100 W (medium treatment). **d,** silver thin film after 5 min O₂ plasma 100 W power (hard treatment). Scale bars, 500 nm.

As a result, the final [AgSePh]∞ films show increased roughness on the hundreds of nanometers scale. This has an obvious impact on the UV-vis spectra of the film that show increased light scattering at all wavelengths (Figure 7a). Figure 7b-d reports SEM images at low magnification of the NC films obtained by different AgO film at increased O₂ plasma conditions: the harder the plasma treatment, the wavier the [AgSePh]∞ NC film.

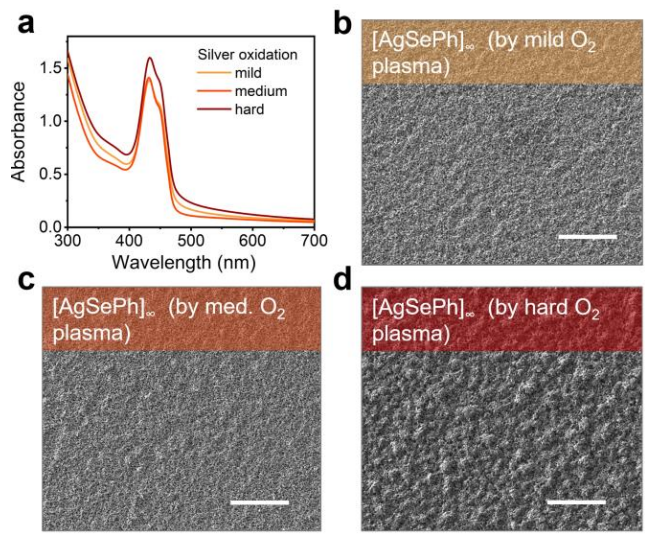

**Figure 7. a,** Ag oxidation influence on [AgSePh]∞ film light absorption. **b-d,** SEM of the [AgSePh]∞ films obtained from silver films that underwent different O₂ plasma treatments. From mild O₂ plasma **(b)** medium O₂ plasma **(c)**, and hard O₂ plasma **(d)**. Scale bars, 5 μm.

Then we turn our focus on the [AgSePh]∞ growth time. At time zero, different sealed vials, containing the silver oxide film and the liquid benzeneselenol, are placed in a static oven at 90 °C. We extract vials from the oven at different times, leaving the samples to cool down before proceeding to the rinsing step. In Figure 8a, UV-vis spectra show a complete disappearance of the AgO broad feature after 30 min reaction. The excitonic peaks show only a little increase as the reaction proceeds past 30 min, indeed the profilometer-measured thickness of the films increases from: 80 nm of the 10 min reaction sample, to about 300 nm of the 30 min reaction sample and then it stays about constant. Nevertheless, the morphology of the film has pronounced variations (Figure 8b-f). The [AgSePh]∞ NCs are already well-shaped after 30 min reaction, but at the 4 h timestamp, something has changed. The crystals appear more clustered, the AFM imaging is more difficult as the surface appears sticky (Supporting Figure S9), possibly indicative of unbound molecular species that are moving on the surface, causing the coalescence of smaller NC into bigger ones, then observed after 24 h growth (Figure 8f). Indeed no silver oxide precursor is left after 30 min (as corroborated by XRD and Raman spectra, Supporting Figure S8) and the possible evolution of NC must then come from a rearrangement of the polymer that seems to be rather mobile at this temperature (90 °C) and conditions (saturated PhSeH atmosphere).

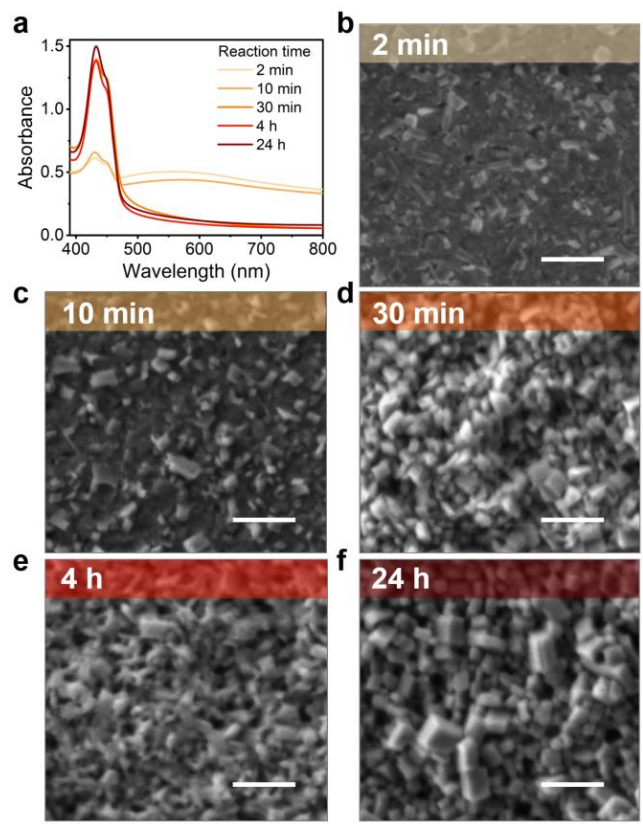

**Figure 8.** Reaction time effect on [AgSePh]∞ film light absorption **(a)** and film morphology **(b-f)**. After 30 min reaction the silver oxide is no longer detected by UV-vis, and the NCs are already formed **(d)**. As the film is left in the reaction environment, the PhSeH acts on the NCs by favoring partial decomposition **(e)**, and subsequent coalescence (**f**) resulting in better-defined, larger NCs. Scale bars, 500 nm.

Finally, the impact of silver film thickness on the growth of the NC film morphology, crystallinity, and optical properties is assessed. Different silver films thicknesses result in silver oxide films having different morphologies and roughness (see AFM in Figure 6b and Supporting Figure S10a,b). The minimum RMS roughness (2.5 nm) is obtained for the 20 nm Ag thin film, which upon mild oxidation double its thickness, while the 5 nm and the 50 nm silver thin films show a little higher RMS roughness (6.4 nm and 8.1 nm, respectively). The thinnest silver film oxidizes resulting in AgO or $Ag_2O$ cubic phases (maybe due to higher surface to volume ratio, pushing the oxidation toward the most stable oxide phase when exposed to air). Thicker Ag films lead to AgO tetragonal phase, with complete silver oxidation, even upon mild $O_2$ plasma (see XRD in Supporting Figure S10c). As previously discussed, this silver oxide phase variation does not lead to a substantial difference in the products in terms of crystallinity. Nevertheless, the oxidized 5 nm silver thin film promotes the growth of peculiar micron-size $[AgSePh]_\infty$ crystals, well aligned parallel to the substrate (Figure 9b). The sample surface is homogenously covered by oriented crystals that give rise to a particular light absorption bump, contributed by 3 excitonic peaks. The latter is very similar to the case of the not purposely oxidized metallic 20 nm silver film (see the discussion in the Supporting Figure S4), which after 4 h of reaction, showed sparse, flat and well surface-aligned $[AgSePh]_\infty$ crystals. This parallel observations lead us to the conclusion that silver oxide ultra-thin films (as it is the case for a native AgO/Ag) promote the reaction of ordered, micron size, silver benzeneselenolate crystals that could be interesting for polarized light detection exploiting the high excitonic anisotropy (in-plane versus out-of-plane) of this material.[9] Thicker Ag films produce instead $[AgSePh]_\infty$ NCs, as already discussed. Based on the SEM images of Figure 9b,d, we note that thicker Ag films lead to larger NCs at fixed oxidation power and reaction time. We speculate that this is a direct consequence of the larger amount of precursors available close to the nucleation spot where the single NC spurs and grows. Moreover it is important to notice that even the 50 nm thick silver film (100 nm AgO after plasma treatment) is completely reacted, leaving no traces of unreacted precursors, while all the peak positions are unchanged respect to different Ag film thicknesses (see XRD, FTIR and Raman spectra reported in Supporting Figure S11).

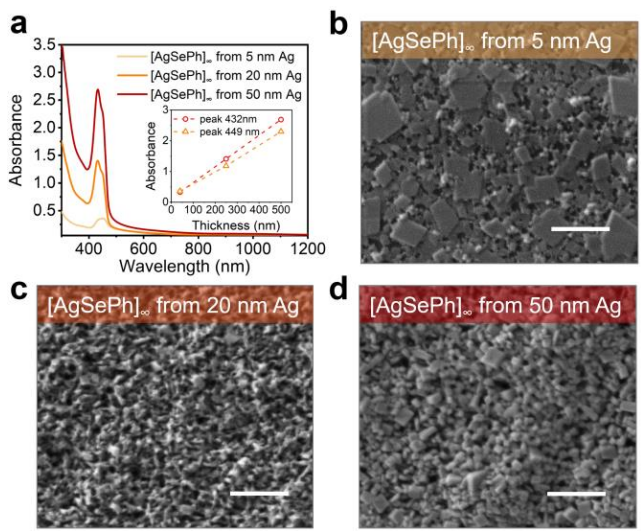

**Figure 9.** Silver film thickness effect on the $[AgSePh]_\infty$ film light absorption (**a**) and morphology (**b-d**). Different Ag thicknesses are tested: 5 nm (**b**), 20 nm (**c**), 50 nm (**d**). The silver is mildly oxidized before the PhSeH reaction and the growth time is held fixed at 4 h. 5 nm thin silver film leads to sparkly different result in the final film, characterized by micron-size flat crystals. Thicker AgO films instead convert to $[AgSePh]_\infty$ films with randomly oriented nanocrystals that influence the excitonic resonances relative weights in the UV-vis experiment (**a**), inset. Scale bars in (**b-d**), 500 nm.

This observation has an impact on the understanding of the low energy absorption tail revealed by optical spectroscopy on every sample hereby presented. In fact, presence of AgO and Ag leftovers could indeed affect the optical absorption spectra by adding broad features to the sharp excitonic peaks. Which were not revealed by a systematic investigation on the silver oxidation effect, benzeneselenol reaction time, and silver film thickness. On the other hand, the scattering contribution to the optical spectra was controlled by changing surface roughness while the distinct low energy tail remained unchanged. Consequently, our analysis suggests a more intrinsic nature this feature to be related to localized in-gap states whose presence seems independent from the many reaction parameters explored in this article.

## Conclusions

While the metallorganic chalcogenides coordination polymers are still in an early development stage, very appealing optoelectronic properties have already been demonstrated. These materials offer the possibility of bridging the gap between ionic lattices of 2D perovskites and the covalent bonds of 2D transition metal dichalcogenides. Despite some earlier works on the synthesis have been produced in the last 20 years, the systematic spectroscopic and electronic investigation of metallorganic chalcogenides polymers have been so far limited by samples morphology. We approached the material's synthesis with a strong focus on the deployment of the polymer $[AgSePh]_\infty$ to optoelectronic devices that typically require contactable continuous films. We demonstrate to reach full control over the reaction process, enabled by a quick in *in-situ* acid-base reaction, leading to homogenous and continuous nanocrystal films. The first vibrational modes assignment for $[AgSePh]_\infty$ is introduced, and FTIR and Raman spectroscopies used for samples purity assessment over possible unreacted precursors. These are excluded by a concerted spectroscopic analysis at X-ray and infrared light. Such observation places a constraint on the origin of the light absorption low-energy tail, important for photoluminescence applications. Our analysis suggests that this intragap states are not related with the material's chemical purity and reinforce the idea that local disorder in the crystal structure, previously observe by TEM *in-situ* experiments, could play a major role in this.[9] The $O_2$ plasma oxidation of silver films is discussed, highlighting the formation of the unstable, mixed valence silver oxide AgO. While the oxide phase is shown not to play a significant role in the reaction process carried out by benzeneselenol, the morphology of the silver oxide film is important. Mild silver oxidations are shown to result in smoother AgO films leading to low-scattering $[AgSePh]_\infty$ NC films. Time-depended reactions experiments showed a quick reagent evolution within tens of minutes. A coalescence phenomenon has been observed on fully reacted $[AgSePh]_\infty$ NC films, driven by PhSeH saturated atmosphere at high temperature. Finally, different silver oxide film thickness were shown to lead to different final product morphologies: micron-size substrate aligned crystals can be of interest as polarized detectors given the strong in- versus out-of- plane exciton anisotropicity. On the other hand, we recently just shown that larger area, homogenous NC films can be used as semiconducting films for photodetection[10] with tunable thickness. Being the silver benzeneselenolate representative of a potentially wide metal organic chalcogenide polymer class, we expect this strategy to be considered for the development of similar materials to develop large-area self-assembled NC films for optoelectronic applications.

## ASSOCIATED CONTENT

The Supporting Information is available free of charge via the Internet at http://pubs.acs.org.

High resolution XPS and detailed spectroscopic (Raman, FTIR) comparison of precursors and products; AgO XRD peaks labelling, and silver oxidation effect on the metallorganic reaction; Silver oxidation XRD, SEM, AFM and UV-vis analysis; Reaction time assessment by XRD, Raman and AFM; Film thickness variation analysis; Vibrational spectroscopy peaks assignment table.

## AUTHORS INFORMATION

### Corresponding Authors

* lorenzo.maserati@iit.it, mario.caironi@iit.it

### Author Contributions

L.M. and M.C. conceived the project. L.M. performed the synthesis, UV-Vis, Raman, FTIR spectrometry, and SEM. M.P. performed the XPS experiments. S.P. performed the X-ray diffraction and topography experiments. The manuscript was written through contributions of all authors. All authors have given approval to the final version of the manuscript.

## ACKNOWLEDGMENTS

L.M. acknowledges the CNST technical staff for their support. L.M. thanks Giorgio Giuffredi for the help in the Raman and FTIR measurements, and Mary S. Collins for the very useful discussions on the reaction chemistry. This work was in part carried out at Polifab, the micro- and nanotechnology center of the Politecnico di Milano.

Work at the CNST was partially financially supported by the European Research Council under the European Union's Horizon 2020 research and innovation program "HEROIC", grant agreement 638059. The work involving the Molecular Foundry was partially supported by the Marie Skłodowska-Curie Research and Innovation Staff Exchange (RISE) project "SONAR", grant agreement 734690.

## ABBREVIATIONS

PhSeH, benzeneselenol; [AgSePh]∞, silver benzeneselenolate coordination polymer; NC, nanocrystal.

TOC Graphic

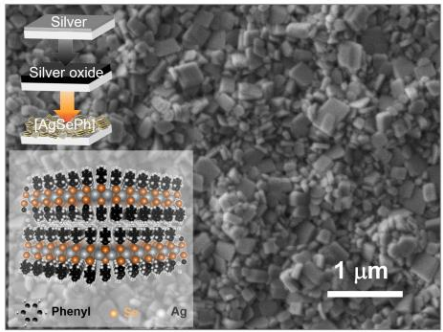